\documentclass[prc,amsmath,amssymb,showpacs,superscriptaddress,twocolumn]
{revtex4}
\usepackage{longtable,dcolumn,epsfig,color}

\begin{document} 

\title{Photoexcitation of $^{76}$Ge}

\author{R.~Schwengner} 
\affiliation{Helmholtz-Zentrum Dresden-Rossendorf, 01328 Dresden, Germany}
\author{R.~Massarczyk}
\affiliation{Los Alamos National Laboratory, Los Alamos, New Mexico 87545, USA}
\author{K.~Schmidt}
\affiliation{Helmholtz-Zentrum Dresden-Rossendorf, 01328 Dresden, Germany}
\author{K.~Zuber}
\affiliation{Institut f\"ur Kern- und Teilchenphysik, Technische Universit\"at
             Dresden, 01069 Dresden, Germany}
\author{R.~Beyer}
\affiliation{Helmholtz-Zentrum Dresden-Rossendorf, 01328 Dresden, Germany}
\author{D.~Bemmerer}
\affiliation{Helmholtz-Zentrum Dresden-Rossendorf, 01328 Dresden, Germany}
\author{S.~Hammer}
\affiliation{Helmholtz-Zentrum Dresden-Rossendorf, 01328 Dresden, Germany}
\author{A.~Hartmann}
\affiliation{Helmholtz-Zentrum Dresden-Rossendorf, 01328 Dresden, Germany}
\author{T.~Hensel}
\affiliation{Helmholtz-Zentrum Dresden-Rossendorf, 01328 Dresden, Germany}
\affiliation{Institut f\"ur Kern- und Teilchenphysik, Technische Universit\"at
             Dresden, 01069 Dresden, Germany}
\author{H.~F.~Hoffmann}
\affiliation{Helmholtz-Zentrum Dresden-Rossendorf, 01328 Dresden, Germany}
\affiliation{Institut f\"ur Kern- und Teilchenphysik, Technische Universit\"at
             Dresden, 01069 Dresden, Germany}
\author{A.~R.~Junghans} 
\affiliation{Helmholtz-Zentrum Dresden-Rossendorf, 01328 Dresden, Germany}
\author{T.~K\"ogler}
\affiliation{Helmholtz-Zentrum Dresden-Rossendorf, 01328 Dresden, Germany}
\affiliation{OncoRay – National Center for Radiation Research in Oncology,
             Faculty of Medicine and University Hospital Carl Gustav Carus,
             Technische Universit\"at Dresden, 01309 Dresden, 
             Helmholtz-Zentrum Dresden-Rossendorf, 01328 Dresden, Germany}
\author{S.~E.~M\"uller}
\affiliation{Helmholtz-Zentrum Dresden-Rossendorf, 01328 Dresden, Germany}
\author{M.~Pichotta}
\affiliation{Institut f\"ur Kern- und Teilchenphysik, Technische Universit\"at
             Dresden, 01069 Dresden, Germany}
\author{S.~Turkat}
\affiliation{Institut f\"ur Kern- und Teilchenphysik, Technische Universit\"at
             Dresden, 01069 Dresden, Germany}
\author{J.~A.~B.~Turko}
\affiliation{Helmholtz-Zentrum Dresden-Rossendorf, 01328 Dresden, Germany}
\affiliation{OncoRay – National Center for Radiation Research in Oncology,
             Faculty of Medicine and University Hospital Carl Gustav Carus,
             Technische Universit\"at Dresden, 01309 Dresden, 
             Helmholtz-Zentrum Dresden-Rossendorf, 01328 Dresden, Germany}
\author{S.~Urla{\ss}} 
\affiliation{Helmholtz-Zentrum Dresden-Rossendorf, 01328 Dresden, Germany}
\affiliation{Conseil Europ{\'e}en pour la Recherche Nucl{\'e}aire (CERN),
             1211 Meyrin, Geneva, Switzerland}
%\affiliation{European Organization for Nuclear Research (CERN), 1211 Meyrin,
%             Geneva, Switzerland}
\author{A.~Wagner} 
\affiliation{Helmholtz-Zentrum Dresden-Rossendorf, 01328 Dresden, Germany}

\date{\today} 

\begin{abstract}
The dipole strength of the nuclide $^{76}$Ge was studied in photon-scattering
experiments using bremsstrahlung produced with electron beams of energies of
7.8 and 12.3 MeV at the $\gamma$ELBE facility. We identified 210 levels up to
an excitation energy of 9.4 MeV and assigned spin $J$ = 1 to most of them.
The quasicontinuum of unresolved transitions was included in the analysis of
the spectra and the intensities of branching transitions were estimated on the
basis of simulations of statistical $\gamma$-ray cascades. The
photoabsorption cross section up to the neutron-separation energy was 
determined and is compared with predictions of the statistical reaction model.
The derived photon strength function is compared with results of experiments
using other reactions.
\end{abstract}


\maketitle

\section{Introduction}
\label{sec:intro}

The search for signals of the neutrinoless double-$\beta$ ($0\nu\beta\beta$)
decay is currently one of the most challenging experimental efforts expected to
gain information about the validity of the standard model of particle physics.
The existence of this decay mode would imply that neutrinos are identical with
antineutrinos, their antiparticles, and that special conditions for their
vertices exist, realized for example through a non-zero neutrino mass
\cite{zub20}. The discovery of this process would prove that the lepton number
is violated by two units and thus, physics goes beyond the standard model.
Experiments searching for this very rare decay mode need large amounts of
target material under very low background conditions. Hence, the experiments
are performed deep underground. Among the nuclides, where $\beta\beta$ decay is
possible in contrast to $\beta$ decay, there is the nuclide $^{76}$Ge, which
can be used at the same time as target and detector material. Collaborations
using this target and detector material are, for example, MAJORANA \cite{alv19}
and GERDA \cite{ago19}. Up to now, GERDA is the experiment that sets the
strongest limit on the half-life of the $0\nu\beta\beta$ decay of $^{76}$Ge to
about $10^{26}$ years \cite{ago18}. The next generation of experiments, such as
LEGEND \cite{abg17}, tries to reach $10^{28}$ years. Thus, lower background
levels have to be reached and every possible background reaction channel has to
be understood as well as possible. Another current project is CDEX-1
\cite{kan13}. If the $0\nu\beta\beta$ decay is realized in nature, the
$\beta$ spectrum is discrete rather than continuous. The signal for this
process is a peak at the $Q$ value of (2039.006 $\pm$ 0.050) keV in the
sum-energy spectrum.

An important issue for this search is the exclusion of signals from other
reactions on $^{76}$Ge that involve the emission of $\gamma$ rays of about this
energy. Earlier experiments using the $^{76}$Ge($n,\gamma$) reaction revealed a
$\gamma$ ray at 2035.5 keV \cite{mei12}. A 2040.7 keV $\gamma$ ray was
identified in the $\beta$ decay of $^{76}$Ga \cite{cam71}. In
$^{76}$Ge($n,n'\gamma$) experiments, $\gamma$ rays at 2034.8 keV \cite{neg13}
and at 2038.9 keV \cite{muk17} were observed. Various experiments using
neutron-induced reactions or activations on Ge isotopes are currently performed
at several laboratories. In addition to these, one may think of other reactions
that can serve to identify so far unknown $\gamma$ transitions in $^{76}$Ge.
One of these reactions is photon scattering $(\gamma,\gamma')$, also called
nuclear resonance fluorescence, in which the incident photons transfer
preferentially angular momentum $L$ = 1 and hence excite states of spin
$J$ = 1 from the ground state in even-even nuclei. In the present work, we
performed photon-scattering experiments at the bremsstrahlung facility
$\gamma$ELBE \cite{sch05} of Helmholtz-Zentrum Dresden-Rossendorf (HZDR) to
study states up to the neutron-separation energy $S_n$ = 9.4 MeV and their
deexcitation to low-lying states, in particular the possible occurrence of
$\gamma$ rays with energies in the interesting region around 2039 keV.

In addition to the just described interest from the search for the
$0\nu\beta\beta$ decay of $^{76}$Ge, photon scattering from $^{76}$Ge is also
of interest for nuclear structure and reaction physics. Photoabsorption cross
sections $\sigma_\gamma$ and the related photon strength functions
$f_1(E_\gamma)$ for $L$ = 1 have attracted growing interest \cite{gor19,kaw20}
because of their importance as inputs to calculations of reaction cross
sections within the statistical reaction model \cite{hau52}. In
photoabsorption, the two quantities are connected via the relation
$\sigma_\gamma = g (\pi \hbar c)^2$ $E_\gamma$ $f_1(E_\gamma)$ with
$g = (2J_x+1)/(2J_0+1)$, where $J_x$ and $J_0$ are the spins of the excited and
ground states, respectively. It was shown, for example, that the so-called
pygmy dipole resonance (PDR) \cite{bar73,sav13,bra19}, an extra strength found
on top of the tail of the isovector giant dipole resonance (GDR), influences
neutron-capture reaction rates \cite{bea12,tso15}, which are important for the
synthesis of heavy elements in astrophysical processes \cite{arn07,kap11}.

In an earlier photon-scattering experiment on $^{76}$Ge using unpolarized
bremsstrahlung at the former Stuttgart Dynamitron and polarized bremsstrahlung
at the former Gie{\ss}en linear accelerator, 30 states with $J$ = 1 were
identified between 2.9 and 9.1 MeV and parities were assigned to 17 of them
\cite{jun95}. Further experiments were performed using bremsstrahlung at the
S-Dalinac electron accelerator of Technische Universit\"at Darmstadt, Germany,
\cite{son11} and quasimonoenergetic, polarized photons at the high-intensity
$\gamma$-ray source (HI$\gamma$S) \cite{wel09} of the Triangle Universities
Nuclear Laboratory (TUNL) in Durham, North Carolina. The experiments are
briefly described in Ref.~\cite{wer15} while 128 states with $1^-$ assignments
and 2 with $1^+$ assignments between 4.4 and 8.9 MeV are compiled in a Ph.D.
thesis \cite{coo15}. In the present work, we found 210 states and assigned
$J$ = 1 to most of them. In addition, we determined the photon-scattering cross
section for 10 keV bins of excitation energy up to $S_n$. In this analysis, the
intensity in the quasi-continuum part of the spectrum was taken into account.
Moreover, we estimated average intensities of inelastic transitions to
low-lying excited states and branching ratios of the ground-state transitions
by means of simulations of statistical $\gamma$-ray cascades. Using these
quantities, we determined the photoabsorption cross section. 

Photon strength functions in $^{76}$Ge have previously been studied using
$\beta$ decay of $^{76}$Ga in connection with the so-called Oslo method
\cite{spy14}. Preliminary results for a photon strength function deduced
from photon-scattering experiments with quasimonoenergetic photons at
HI$\gamma$S were presented in Ref.~\cite{ton17}.

\section{Experimental Methods and results}
\label{sec:exp}

\subsection{The photon-scattering method}
\label{sec:meth}

In photon-scattering experiments, the energy- and solid-angle-integrated
scattering cross section $I_s$ of an excited state at the energy $E_x$ is
deduced from the measured intensity of the respective transition to the ground
state. It can be determined relative to known integrated scattering cross
sections. In the present experiments, we used the integrated scattering cross
sections $I_s(E_x^{\rm B})$ of states in $^{11}$B \cite{ajz90} and their
angular correlations including mixing ratios \cite{rus09B} as a reference:

\begin{eqnarray}
\label{eq:sigs}
\frac{I_s(E_x)}{I_s(E_x^{\rm B})} = 
\left(\frac{I_\gamma(E_\gamma,\theta)} {W(E_\gamma,\theta)
  \Phi_\gamma(E_x) N_N}\right) \nonumber \\
  \times \left(\frac{I_\gamma(E_\gamma^{\rm B},\theta)}
{W(E_\gamma^{\rm B},\theta) \Phi_\gamma(E_x^{\rm B}) N_N^{\rm B}}\right)^{-1}.
\end{eqnarray}

Here, $I_\gamma(E_\gamma,\theta)$ and $I_\gamma(E_\gamma^{\rm B},\theta)$
denote the efficiency-corrected measured intensities of a considered
ground-state transition at energy $E_\gamma$ and of a ground-state transition in
$^{11}$B at $E_\gamma^{\rm B}$, respectively, observed at an angle $\theta$ to 
the beam. $W(E_\gamma,\theta)$ and $W(E_\gamma^{\rm B},\theta)$ describe the
angular correlations of these transitions. The quantities $N_N$ and
$N_N^{\rm B}$ are the areal densities of nuclei in the $^{76}$Ge and $^{11}$B
targets, respectively. The quantities $\Phi_\gamma(E_x)$ and 
$\Phi_\gamma(E_x^{\rm B})$ stand for the photon fluxes at the energy of the
considered level and at the energy of a level in $^{11}$B, respectively.

The integrated scattering cross section is related to the partial width of the
ground-state transition $\Gamma_0$ according to
\begin{equation}
\label{eq:gam}
I_s = \int\limits^{+\infty}_{0} \sigma_{\gamma \gamma} ~dE =
            \left(\frac{\pi \hbar c}{E_x}\right)^2
            \frac{2 J_x + 1}{2 J_0 + 1}
            \frac{\Gamma_0^2}{\Gamma},
\end{equation}
where $\sigma_{\gamma \gamma}$ is the elastic scattering cross section, $E_x$,
$J_x$ and $\Gamma$ denote energy, spin and total width of the excited level,
respectively, and $J_0$ is the spin of the ground state. If a given level
deexcites to low-lying excited states (inelastic scattering) in addition to the
deexcitation to the ground state (elastic scattering), then the branching ratio
$b_0 = \Gamma_0/\Gamma$ of the ground-state transition has to be known to deduce
$\Gamma_0$. The $\gamma$-ray intensities and, hence, the deduced quantities
$I_s$ and $\Gamma_0$ are also distorted if a level is populated from
higher-lying levels. This feeding can be reduced by choosing beam energies not
far above the considered levels.

Spins of excited states are deduced by comparing experimental ratios of
$\gamma$-ray intensities, measured at two angles, with theoretical predictions.
The optimum combination includes angles of 90$^\circ$ and 127$^\circ$ to the
beam direction, because the respective ratios for the spin sequences $0-1-0$
and $0-2-0$ differ most at these angles. The expected values are
$W(90^\circ)/W(127^\circ)_{0-1-0} = 0.74$ and
$W(90^\circ)/W(127^\circ)_{0-2-0} = 2.15$ by taking into account opening
angles of $16^\circ$ and $14^\circ$ of the collimators in front of the detectors
placed at $90^\circ$ and $127^\circ$, respectively, in the setup at
$\gamma$ELBE \cite{sch05}.

\subsection{The target}
\label{sec:target}

The target consisted of 1.8760 g of germanium, enriched to 93.4\%  in $^{76}$Ge,
in a square shape of 15 mm $\times$ 11 mm. The germanium target was combined
with 0.300 g of boron, enriched to 99.5\% in $^{11}$B and formed to a disk of
20 mm in diameter. The known integrated scattering cross sections of levels in
$^{11}$B were used to determine the photon flux (see Sec.~\ref{sec:expelbe}).
The photon-flux density was proven to be nearly constant in a spot of about
25 mm in diameter \cite{rus07}. For the calculation of cross sections for
$^{76}$Ge, the ratio of the $^{76}$Ge and $^{11}$B target areas was taken into
account.

\subsection{Detector response}
\label{sec:effflux}

For the determination of the integrated scattering cross sections according
to Eq.~(\ref{eq:sigs}) the relative efficiencies of the detectors and the
photon flux are needed. The detector response was simulated using the program
package GEANT4 \cite{ago03,all06,all16}. The reliability of the simulations was
tested by comparing simulated spectra with measured ones as described, for
example, in Refs.~\cite{rus07,sch07,rus08,mas12,mar10}. The determination of
the absorption cross section requires in addition a correction of the
experimental spectra for photons scattered by atomic processes induced by the
impinging photons in the target material, and for ambient background radiation,
which is described in Sec.~\ref{sec:corr}.

The absolute efficiencies of the high-purity germanium (HPGe) detectors in the
setup at $\gamma$ELBE were determined experimentally up to 2.4 MeV from
measurements with a $^{226}$Ra calibration source. For interpolation, an
efficiency curve calculated with GEANT4 and scaled to the absolute experimental
values was used. A check of the simulated efficiency curve up to about 9 MeV
was performed via various $(p,\gamma)$ reactions at the HZDR Tandetron
accelerator. The efficiency values deduced from these measurements agree with
the simulated values within their uncertainties \cite{tro09}. Similar results
were obtained for the resonances at 4.44 and 11.66 MeV in $^{12}$C populated in
the $^{11}$B$(p,\gamma)$ reaction at the van-de-Graaff accelerator of the
Triangle Universities Nuclear Laboratory (TUNL) in Durham, North Carolina
\cite{car10}.

\subsection{Experiments and results}
\label{sec:expelbe}

\begin{figure}
\epsfig{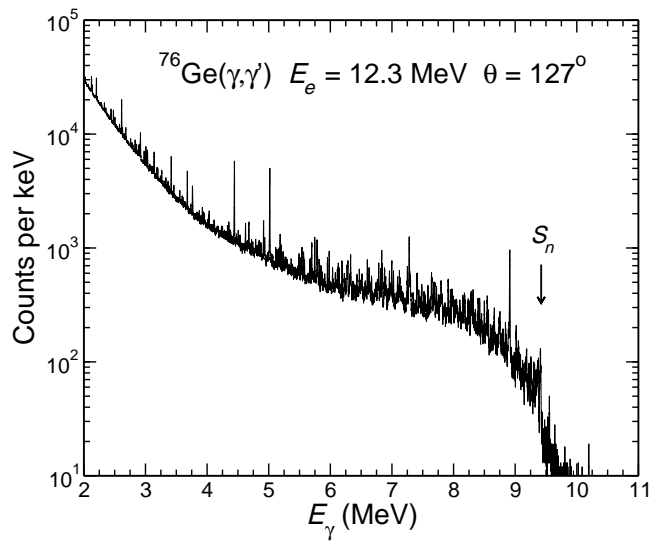}
  \caption{\label{fig:elbespec} Part of a spectrum of photons scattered from
    $^{76}$Ge combined with $^{11}$B, measured during the irradiation with
    bremsstrahlung produced by electrons of an energy of $E_e^{\rm kin}$ =
    12.3 MeV. This spectrum is the sum of the spectra measured with the two
    detectors placed at 127$^\circ$ relative to the beam at $\gamma$ELBE. The
    arrow labeled $S_n$ indicates the neutron-separation energy.}
\end{figure}

The nuclide $^{76}$Ge was studied in two experiments at $\gamma$ELBE
\cite{sch05}. Bremsstrahlung was produced using electron beams of 7.8 and 12.3
MeV kinetic energy, respectively. In the measurement at 7.8 MeV, the electron
beam hit a niobium foil of 7 $\mu$m thickness acting as a radiator at an
average current of about 700 $\mu$A. In the measurement at 12.3 MeV, the
niobium foil had a thickness of 12.5 $\mu$m and the average current was also
about 700 $\mu$A. A 10 cm thick aluminum absorber (beam hardener) was placed
behind the radiator to reduce the low-energy intensity of the bremsstrahlung
spectrum in the measurement at 12.3 MeV. The photon beam was collimated by a
260-cm-long pure-aluminum collimator with a conical borehole of 8 mm in
diameter at the entrance, 90 cm behind the radiator, and 24 mm in diameter at
the exit. The target, placed 200 cm behind the collimator exit, was irradiated
with a typical flux of about $10^9$ s$^{-1}$ in a spot of 38 mm in diameter.
Scattered photons were measured with four HPGe detectors with a full-energy
efficiency of 100\% relative to a NaI detector of 7.6 cm in diameter and 7.6 cm
in length. All HPGe detectors were surrounded by escape-suppression shields
made of bismuth germanate (BGO) scintillation detectors of 3 cm in thickness.
Two HPGe detectors were placed vertically at 127$^\circ$ relative to the
photon-beam direction and a distance of 32 cm from the target. The other two
HPGe detectors were positioned in a horizontal plane at 90$^\circ$ to the beam
and a distance of 28 cm from the target. Absorbers of 8 mm Pb plus 3 mm Cu were
placed in front of the detectors at 90$^\circ$ and of 3 mm Pb plus 3 mm Cu in
front of the detectors at 127$^\circ$. Spectra of scattered photons were
measured for 132 h each in the experiments at 7.8 and 12.3 MeV electron energy.
Part of a spectrum including events measured with the two detectors placed at
127$^\circ$ relative to the beam at an electron energy of 12.3 MeV is shown in
Fig.~\ref{fig:elbespec}.  

\begin{figure}
\epsfig{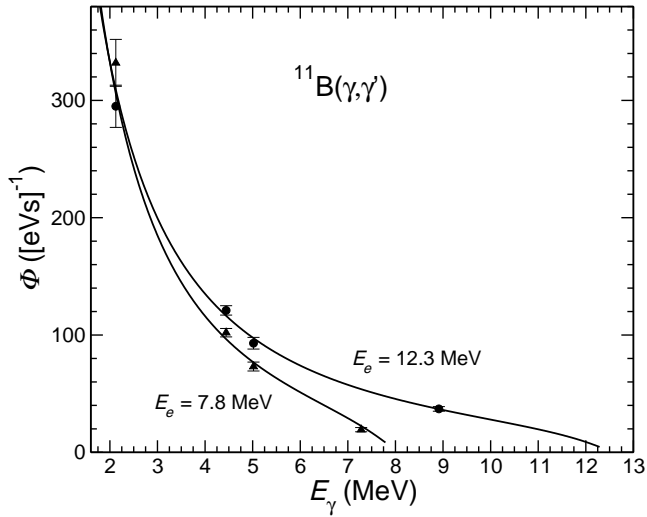}
\caption{\label{fig:flux} Average absolute photon flux on the $^{11}$B target
  deduced from intensities of known transitions in $^{11}$B for the
  measurements at $E_e$ = 7.8 MeV (triangles) and $E_e$ = 12.3 MeV (circles) at
  $\gamma$ELBE. The curves represent the calculated flux described in the
  text. }
\end{figure}

The absolute photon fluxes in the two measurements at $\gamma$ELBE were
determined from intensities and known integrated scattering cross sections of
transitions in $^{11}$B. The 7283 keV transition in $^{11}$B was found to form
an unresolved doublet with another transition, which is negligible at
$E_e$ = 7.8 MeV, but comparable in intensity with the transition in $^{11}$B at
$E_e$ = 12.3 MeV. The latter has therefore not been considered for the flux
determination. For interpolation, the photon flux was calculated using a
bremsstrahlung computer code \cite{hau05} based on the Born approximation with
Coulomb correction \cite{roc72} and including an atomic screening correction
\cite{sal87}. In addition, the flux was corrected for the attenuation by the
beam hardener by applying a parametrization of the results of a corresponding
GEANT4 simulation. The calculated flux curves were adjusted to the experimental
values obtained at the energies of levels in $^{11}$B. The experimental flux
values and the calculated curves are presented in Fig.~\ref{fig:flux}.

The measurements at two electron energies allowed us to identify inelastic
transitions that feed low-lying from high-lying levels. Transitions found in the
measurement at $E_e^{\rm kin}$ = 7.8 MeV are assumed to be ground-state
transitions. Additional transitions observed up to 7.8 MeV in the measurement
at $E_e^{\rm kin}$ = 12.3 MeV are considered to be inelastic transitions from
high-lying to low-lying excited states. By comparing the respective spectra,
these inelastic transitions were sorted out. Besides, there is a number of
transitions with energies that fit the difference between the energy of a
higher-lying level and the first or second excited $2^+$ states. These
transitions are also assumed to be inelastic transitions, if their intensity is
smaller than that of the ground-state transition from the considered
higher-lying level. The remaining ground-state transitions were used to derive
the corresponding level energies, the integrated scattering cross sections of
the states, and spin assignments deduced from angular distributions of the
ground-state transitions. These quantities are compiled in Table~\ref{tab:gam}.
The integrated scattering cross sections of levels up to $E_x$ = 7.0 MeV were
taken from the measurement at 7.8 MeV electron energy, because they are
affected by feeding intensities in the 12.3 MeV measurement. We note that in
principle low-lying states can also be fed from other states below 7.8 MeV.
However, previous investigations have shown that the states below about
6 MeV are fed by states mainly above about 9 MeV \cite{sch08,ben09}.

\begin{longtable}{cccc}
  \caption{\label{tab:gam}Levels assigned to $^{76}$Ge.} \\
  \hline
  \hline
  $E_x$(keV)\footnotemark[1] &
  $I_\gamma(90^\circ)/I_\gamma(127^\circ)$\footnotemark[2] &
  $J_x^\pi$\footnotemark[3]  &
  $I_s$(eV b)\footnotemark[4] \\
  \hline                                      
    564.5(1)  &  1.05(14) & $2^+$\footnotemark[5] & \\
   1109.2(1)  &  1.11(15) & $2^+$\footnotemark[5] & \\
   2504.2(4)  &           & $2^+$\footnotemark[5] & \\
   2655.1(3)  &  0.79(27) & (1) &    5.6(10)  \\
   2919.3(2)  &  1.33(24) & $1^+$\footnotemark[5] & 12.1(12)    \\
   3006.7(2)  &  0.90(16) & $1^+$\footnotemark[5] &  9.6(10)    \\
   3140.9(2)  &  1.05(15) & $1^+$\footnotemark[5] & 12.0(12)    \\
   3200.0(2)  &  1.23(23) &     &    9.7(11)  \\
   3418.9(1)  &  0.86(4)  & $1^+$\footnotemark[6] &   46(4)     \\
   3680.4(1)  &  0.84(3)  & $1^-$\footnotemark[6] &   52(4)     \\
   3763.2(1)  &  0.83(6)  & $1^+$\footnotemark[6] &   25.8(24)  \\
   3951.0(4)  &  1.10(28) &     &    4.5(7)   \\
   4024.0(2)  &  0.70(20) &  $1^{(-)}$\footnotemark[6] &  6.2(8) \\
   4035.0(2)  &  0.76(15) &  1  &    9.1(10)  \\
   4115.9(2)  &  0.81(12) &  1  &   14.7(16)  \\
   4250.8(3)  &  0.68(20) &  1  &    6.7(11)  \\
   4624.0(2)  &  0.84(8)  & $1^+$\footnotemark[6] &   26.9(24)  \\
   4661.0(4)  &  0.53(16) &  1  &    6.4(10)  \\
   4678.1(1)  &  0.75(5)  &  1  &   33.0(28)  \\
   4722.2(2)  &  0.88(15) & (1) &    9.6(12)  \\
   4741.0(2)  &  0.94(15) &     &   10.3(12)  \\
   4788.9(3)  &  1.17(26) &     &    7.1(11)  \\
   4837.0(4)  &  0.88(20) & (1) &    7.3(12)  \\
   4845.9(3)  &  0.76(13) & 1   &   10.9(13)  \\
   4874.5(2)  &  1.5(6)   &     &    8.7(13)  \\
   4916.5(1)  &  0.75(4)  &  1  &   50(4)     \\
   4935.9(2)  &  0.91(7)  &  1  &   19.5(18)  \\
   5116.4(2)  &  0.85(9)  &  1  &   17.3(18)  \\
   5166.7(2)  &  0.91(9)  & (1) &   17.3(16)  \\
   5185.8(1)  &  0.98(6)  & (1) &   38(3)     \\
   5202.3(2)  &  0.80(8)  &  1  &   22.3(20)  \\
   5222.0(3)  &  1.04(15) &     &   13.3(16)  \\
   5266.8(3)  &  0.65(13) &  1  &   11.6(16)  \\
   5273.6(6)  &  0.80(24) & (1) &    5.0(10)  \\
   5284.9(2)  &  0.60(10) &  1  &   13.1(15)  \\
   5304.1(3)  &  0.56(13) &  1  &   10.2(13)  \\
   5365.6(3)  &  0.68(12) &  1  &   10.1(13)  \\
   5379.5(4)  &  0.59(16) &  1  &    6.2(10)  \\
   5390.6(5)  &  0.81(22) & (1) &    5.4(10)  \\
   5418.6(4)  &  0.76(27) & (1) &    5.4(10)  \\
   5434.3(3)  &  0.74(22) &  1  &    7.9(11)  \\
   5492.7(2)  &  0.62(13) &  1  &   21.3(25)  \\
   5540.1(2)  &  0.76(5)  &  1  &   31.6(27)  \\
   5567.4(2)  &  0.92(8)  & (1) &   20.6(19)  \\
   5581.0(2)  &  0.66(8)  &  1  &   16.0(16)  \\
   5665.2(3)  &  0.72(10) &  1  &   20.8(24)  \\
   5677.6(3)  &  0.75(11) &  1  &   23.5(26)  \\
   5698.8(2)  &  0.74(6)  & $1^-$\footnotemark[6] &   52(5)     \\
   5708.4(6)  &  0.98(21) & (1) &    8.3(14)  \\
   5748.3(1)  &  0.81(5)  & $1^-$\footnotemark[6] &   67(5)     \\
   5785.0(2)  &  0.72(5)  &  1  &   52(4)     \\
   5794.1(2)  &  0.66(7)  &  1  &   26.5(25)  \\
   5820.8(6)  &           &     &   22(5)     \\
   5825.3(8)  &  0.76(10) &  1  &   12(3)     \\
   5846.5(7)  &  0.81(28) &     &    8.8(21)  \\
   5864.8(6)  &  0.83(22) &     &   10.8(22)  \\
   5908.8(3)  &           &     &    8.1(13)  \\
   5954.8(2)  &  0.71(5)  &  1  &   41(3)     \\
   5983.0(2)  &  0.69(4)  & $1^-$\footnotemark[6] &   49(4)     \\
   6048.4(4)  &  0.81(20) &  1  &    7.8(13)  \\
   6081.4(4)  &  0.90(16) & (1) &   12.4(17)  \\
   6102.0(9)  &           &     &    5.3(12)  \\
   6113.6(3)  &  0.77(9)  &  1  &   21.1(22)  \\
   6130.3(2)  &  0.76(7)  &  1  &   30.9(29)  \\
   6145.6(2)  &  0.77(9)  &  1  &   24.2(25)  \\
   6162.4(9)  &  0.9(4)   &     &    4.5(14)  \\
   6191.3(2)  &  0.72(6)  &  1  &   41(4)     \\
   6223.4(7)  &           &     &    6.0(14)  \\
   6228.2(4)  &  0.79(11) &  1  &   14.5(20)  \\
   6234.8(9)  &           &     &    4.6(12)  \\
   6240.7(3)  &  0.70(14) &  1  &   13.3(18)  \\
   6272.7(3)  &  0.79(9)  &  1  &   24.0(25)  \\
   6285.3(2)  &  0.68(6)  &  1  &   46(4)     \\
   6315.4(4)  &  0.66(11) &  1  &   21.5(26)  \\
   6330.2(2)  &  0.77(5)  &  1  &   63(5)     \\
   6366.2(11) &  1.0(3)   &     &    8.7(21)  \\
   6393.2(5)  &  0.81(19) &  1  &   15.5(26)  \\
   6408.1(5)  &  0.63(29) &  1  &   17(3)     \\
   6436.1(9)  &  1.2(4)   &     &   12.0(28)  \\
   6448.3(11) &  0.9(4)   &     &    8.2(23)  \\
   6472.2(3)  &  0.79(9)  &  1  &   41(4)     \\
   6497.9(3)  &  0.72(13) &  1  &   18.7(27)  \\
   6513.3(4)  &  0.75(11) &  1  &   31(4)     \\
   6572.0(6)  &  1.08(21) &     &   11.8(18)  \\
   6601.2(2)  &  0.77(6)  &  1  &   53(5)     \\
   6611.1(6)  &           &     &   13.0(19)  \\
   6629.0(3)  &  0.79(9)  &  1  &   27.0(28)  \\
   6641.9(5)  &  1.00(15) &     &   16.5(21)  \\
   6661.4(9)  &           &     &    9.6(19)  \\
   6670.6(3)  &  0.85(9)  &  1  &   33(3)     \\
   6741.6(6)  &  0.86(20) & (1) &   11.4(19)  \\
   6764.8(4)  &  0.74(11) &  1  &   20.4(24)  \\
   6786.7(2)  &  0.83(6)  &  1  &   59(5)     \\
   6816.5(3)  &  0.75(7)  &  1  &   44(4)     \\
   6835.5(2)  &  0.70(5)  &  1  &   82(7)     \\
   6846.2(3)  &  0.66(7)  &  1  &   44(4)     \\
   6880.3(4)  &  0.70(6)  &  1  &   46(5)     \\
   6884.2(10) &           &     &   18(2)     \\
   6898.9(5)  &  0.55(8)  &  1  &   41(6)     \\
   6908.0(18) &  1.2(3)   &     &   19(5)     \\
   6938.6(7)  &  0.61(16) &  1  &   15.1(27)  \\
   6959.9(3)  &  0.71(8)  &  1  &   45(5)     \\
   6985.1(5)  &  0.53(9)  &  1  &   26(3)     \\
   6998.7(3)  &  0.68(5)  & $1^-$\footnotemark[6] &   79(7)     \\
   7011.0(9)  &  0.46(16) &  1  &   12.3(25)  \\
   7025.8(3)  &  0.73(8)  & $1^{(-)}$\footnotemark[6] & 51(5)    \\
   7047.9(9)  &  0.58(22) &  1  &   10.4(26)  \\
   7081.2(9)  &  0.53(29) &  1  &    9.0(22)  \\
   7091.4(4)  &  0.74(11) &  1  &   28(3)     \\
   7102.4(6)  &  0.75(26) &  1  &   13.8(24)  \\
   7121.3(3)  &  0.80(8)  &  1  &   42(4)     \\
   7130.1(3)  &  0.70(9)  &  1  &   36(4)     \\
   7147.3(4)  &  0.76(13) &  1  &   23(3)     \\
   7171.6(9)  &           &     &   10.8(26)  \\
   7250.5(2)  &  0.79(8)  & $1^-$\footnotemark[6] &   76(7)     \\
   7289.7(4)  &  0.9(5)   &     &   51(5)     \\
   7300.7(3)  &  0.75(9)  & $1^-$\footnotemark[6] &   56(5)     \\
   7406.7(3)  &  0.76(11) &  1  &   60(6)     \\
   7415.6(4)  &  0.98(18) &     &   37(5)     \\
   7452.2(5)  &           &     &   24(4)     \\
   7478.6(5)  &  0.8(3)   &     &   21(3)     \\
   7485.0(3)  &  0.65(26) &  1  &   33(4)     \\
   7521.2(5)  &  0.75(14) &  1  &   26(4)     \\
   7536.6(4)  &  0.89(15) & (1) &   30(4)     \\
   7548.8(7)  &  0.95(26) & (1) &   19(4)     \\
   7584.6(4)  &  0.67(9)  &  1  &   46(5)     \\
   7642.6(4)  &  0.76(9)  &  1  &   69(7)     \\
   7650.8(4)  &  0.79(10) &  1  &   58(6)     \\
   7677.7(4)  &  0.78(12) &  1  &   44(5)     \\
   7694.0(3)  &  0.75(10) &  1  &   59(7)     \\
   7722.7(4)  &  0.93(19) & (1) &   36(6)     \\
   7776.9(7)  &  0.83(17) & (1) &   28(5)     \\
   7783.8(9)  &           &     &   24(5)     \\
   7796.6(4)  &  0.70(10) &  1  &   66(7)     \\
   7803.7(6)  &  0.76(13) &  1  &   42(5)     \\
   7814.3(7)  &  0.79(16) &  1  &   26(4)     \\
   7817.2(2)  &           &     &   20.7(17)  \\
   7836.3(6)  &           &     &   17(4)     \\
   7849.3(5)  &  0.77(24) & (1) &   22(4)     \\
   7861.2(4)  &  0.51(20) &  1  &   31(6)     \\
   7883.3(10) &  0.69(26) &  1  &   16(4)     \\
   7893.6(12) &  1.0(4)   &     &   11(3)     \\
   7913.4(2)  &  0.80(7)  &  1  &   85(8)     \\
   7949.9(2)  &  0.86(12) &  1  &   42(5)     \\
   7975.6(7)  &  0.93(16) & (1) &   29(5)     \\
   7995.8(4)  &  0.85(18) & (1) &   25(4)     \\
   8017.5(14) &  0.7(3)   & (1) &   19(5)     \\
   8026.5(8)  &  0.86(25) & (1) &   34(7)     \\
   8049.3(6)  &  0.82(21) & (1) &   34(5)     \\
   8063.4(8)  &  0.71(27) &  1  &   20(4)     \\
   8094.2(8)  &           &     &   12(4)     \\
   8102.8(5)  &  1.23(28) &     &   24(5)     \\
   8109.5(8)  &           &     &   13(3)     \\
   8134.5(11) &  1.4(5)   &     &   13(4)     \\
   8151.6(5)  &  0.85(12) & $1^{(-)}$\footnotemark[6] &  52(7)   \\
   8160.2(9)  &           &     &   25(5)     \\
   8177.8(4)  &  0.70(8)  &  1  &   56(6)     \\
   8187.8(5)  &  0.86(11) &  1  &   49(6)     \\
   8236.4(4)  &  0.90(14) & (1) &   44(6)     \\
   8252.9(9)  &           &     &   25(6)     \\
   8259.6(6)  &  0.86(16) & (1) &   43(7)     \\
   8284.5(3)  &  0.88(13) & (1) &   72(8)     \\
   8294.3(12) &  0.86(25) &     &   21(4)     \\
   8303.5(5)  &  0.77(13) &  1  &   49(6)     \\
   8317.8(3)  &  0.74(9)  &  1  &   77(7)     \\
   8328.9(7)  &  0.77(15) &  1  &   27(4)     \\
   8347.7(9)  &  0.84(25) &     &   19(4)     \\
   8357.4(7)  &  0.91(29) & (1) &   28(5)     \\
   8397.3(5)  &           &     &   38(6)     \\
   8418.0(15) &           &     &    9(3)     \\
   8425.2(3)  &  1.18(26) &     &   53(7)     \\
   8446.1(7)  &  0.5(4)   &  1  &  16(3)      \\
   8461.9(9)  &           &     &   13(3)     \\
   8500.0(3)  &  0.84(11) &  1  &   62(7)     \\
   8520.7(6)  &           &     &   23(6)     \\
   8535.1(5)  &  0.61(20) &  1  &   33(5)     \\
   8546.1(5)  &  0.68(11) & $1^-$\footnotemark[6] &   73(10)    \\
   8552.3(8)  &  0.42(20) &  1  &   34(7)     \\
   8566.9(3)  &  0.78(12) &  1  &   49(6)     \\
   8602.3(5)  &           &     &   19(4)     \\
   8625.7(7)  &  0.49(25) &  1  &   28(11)    \\
   8649.1(8)  &  1.20(28) &     &   22(4)     \\
   8662.0(4)  &  0.89(12) & (1) &   52(6)     \\
   8696.2(7)  &           &     &   20(10)    \\
   8740.7(4)  &  0.92(13) & (1) &   41(5)     \\
   8752.8(4)  &  1.05(14) & $1^-$\footnotemark[6] &   43(5)     \\
   8768.4(9)  &  0.62(25) &  1  &   15(4)     \\
   8806.2(5)  &  1.2(5)   &     &   27(7)     \\
   8843.7(4)  &  0.67(11) &  1  &   39(6)     \\
   8888.5(9)  &           &     &   19(5)     \\
   9014.2(14) &           & $1^-$\footnotemark[6] &   24(9)     \\
   9019.5(10) &  0.98(15) & (1) &   37(12)    \\
   9033.1(9)  &  1.0(3)   &     &   15(4)     \\
   9051.7(12) &  0.89(17) & (1) &   17(5)     \\
   9058.5(11) &  0.92(29) &     &   18(5)     \\
   9163.3(9)  &  0.69(23) &  1  &   13.9(28)  \\
   9175.5(8)  &  0.59(16) &  1  &   19.1(29)  \\
   9187.4(4)  &  0.80(12) &  1  &   35(4)     \\
   9254.6(7)  &  1.08(29) &     &   21(4)     \\
   9264.1(6)  &  1.1(3)   &     &   22(3)     \\
   9305.0(4)  &           &     &   18.0(26)  \\
   9315.8(4)  &  1.0(5)   &     &   23(3)     \\
   9337.8(6)  &  1.0(3)   &     &   17(3)     \\
   9354.5(8)  &  0.92(29) & (1) &   15.4(27)  \\
   9365.9(5)  &  0.81(18) &  1  &   29(4)     \\
   9377.9(4)  &  0.95(18) & (1) &   38(5)     \\
   9399.4(6)  &  0.60(17) &  1  &   19(3)     \\
   9409.9(4)  &  0.69(11) &  1  &   49(5)     \\
   9417.6(5)  &  0.52(13) &  1  &   29(4)     \\
   9556.6(5)  &  0.71(22) &  1  &   11.5(22)  \\
\hline
\hline
\footnotetext[1]{
Excitation energy. The uncertainty of this and the other quantities in the
table is given in parentheses in units of the last digit. This energy value was
deduced from the $\gamma$-ray energy measured at 127$^\circ$ including a recoil
and Doppler-shift correction.}
\footnotetext[2]{
Ratio of the intensities measured at angles of 90$^\circ$ and 127$^\circ$.
The expected values for an elastic dipole transition (spin sequence $0-1-0$)
and for an elastic quadrupole transition (spin sequence $0-2-0$) are 0.74
and 2.15, respectively.}
\footnotetext[3]{
  Spin deduced from the angular distribution of the ground-state transition.
  A tentative assignment of (1) is given, if the angular distribution is
  compatible with dipole as well as isotropic behavior.}
\footnotetext[4]{
Energy-integrated scattering cross section. Below an excitation energy of
7.0 MeV the value was deduced from the measurement at 7.8 MeV electron energy,
otherwise the value was deduced from the measurement at 12.3 MeV.}
\footnotetext[5]{
Spin and parity taken from Ref.~\cite{muk17}.}
\footnotetext[6]{
Spin and parity taken from Ref.~\cite{jun95}.}
\end{longtable}

\begin{figure}
\epsfig{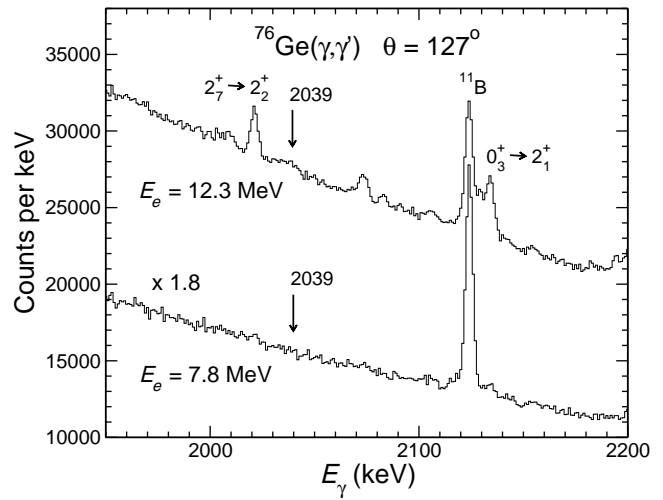}
\caption{\label{fig:2039} Parts of spectra measured at $E_e$ = 7.8 MeV (bottom)
  and $E_e$ = 12.3 MeV (top) showing the energy section around the expected
  2039 keV signal. Note that the 7.8 MeV spectrum was scaled up by a factor of
  1.8 to make the two spectra better comparable. The labels
  $2^+_7 \rightarrow 2^+_2$ and $0^+_3 \rightarrow 2^+_1$ mark the corresponding
  transitions in $^{76}$Ge, 2039 is the energy of the $0\nu\beta\beta$ signal
  and $^{11}$B denotes a transition in this nuclide.}
\end{figure}

A transition with an energy around 2039 keV, close to the one of the expected
signal for the $0\nu\beta\beta$ decay, has not been clearly identified in the
present experiments. In particular, the ground-state transition from the
3951 keV level, known from $^{76}$Ga $\beta$ decay \cite{cam71}, has been
detected in the present study (see Table~\ref{tab:gam}), but there is no
indication of the 2040 keV transition depopulating the 3951 to the 1911 keV
level in the measurement at $E_e$ = 7.8 MeV. There may be a tiny bump in
the measurement at $E_e$ = 12.3 MeV, which occurs due to feeding of the
3951 keV level from higher-lying levels. This situation is illustrated in
Fig.~\ref{fig:2039}. The tentative spin and parity assignment of $(1,2^+)$
given in Ref.~\cite{cam71} for the 3951 keV level could not be made more
precise because of the uncertain angular distribution of the weak ground-state
transition. The spectrum at 12.3 MeV contains, for example, transitions from
the $0^+_3$ state at 2697 and from the $2^+_7$ state at 3130 keV, which are not
seen in the spectrum at 7.8 MeV. This proves that these low-lying states are
mainly fed by states above about 8 MeV.

\section{Determination of the dipole-strength distribution}
\label{sec:corr}

The determination of the dipole-strength distribution and the related
photoabsorption cross section requires the knowledge of the intensity
distribution of the ground-state transitions and their branching ratios. As
these cannot be derived directly from the measured spectra, we applied
statistical methods to discriminate between $\gamma$~rays from nuclear
excitations and photons scattered by atomic processes, and to disentangle the
intensity distributions of elastic and inelastic transitions in the
quasicontinuum of nuclear levels.

First, a spectrum of the ambient background adjusted to the intensities of
the transitions from $^{40}$K and $^{208}$Tl decay in the in-beam spectrum was
subtracted from the measured spectrum. To correct the measured spectrum for the 
detector response, spectra of monoenergetic $\gamma$ rays were calculated in
steps of 10 keV by using the simulation code GEANT4. Starting from the
high-energy end of the experimental spectrum, the simulated spectra were
subtracted sequentially (spectrum-stripping method \cite{fur68}).

\begin{figure}
\epsfig{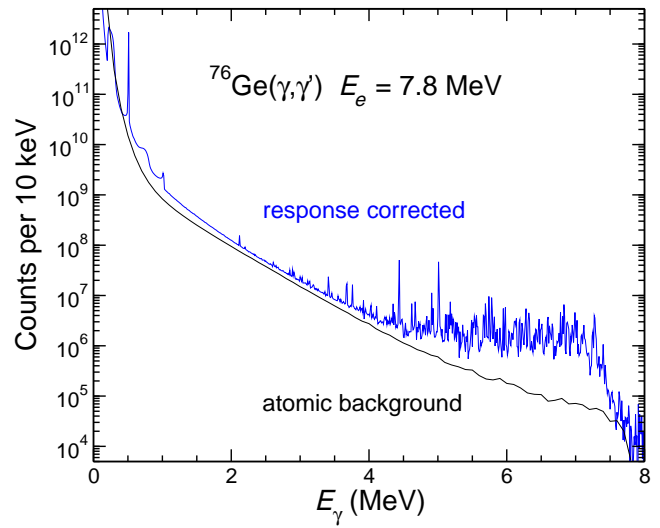}
\caption{\label{fig:AtBgle}Spectrum of the two detectors at 127$^\circ$ and
  $E_e$ = 7.8 MeV, corrected for detector response (blue), and simulated
  spectrum of photons scattered from the target to the detectors
  by atomic processes (black).}
\end{figure}

\begin{figure}
\epsfig{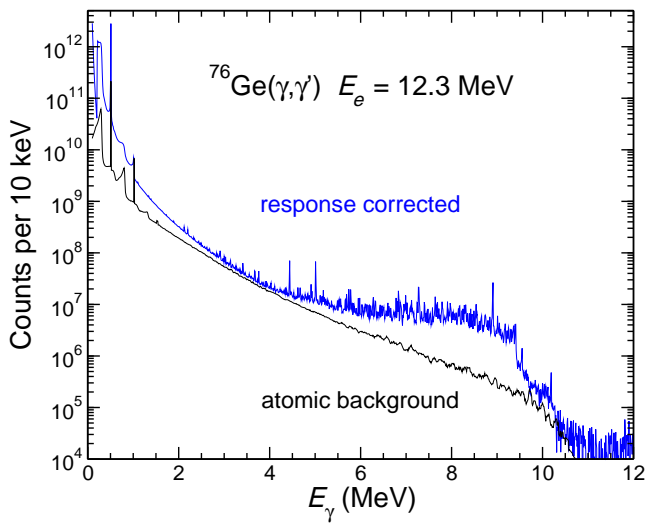}
\caption{\label{fig:AtBghe}Spectrum of the two detectors at 127$^\circ$ and
  $E_e$ = 12.3 MeV corrected for detector response (blue), and simulated
  spectrum of photons scattered from the target to the detectors
  by atomic processes (black).}
\end{figure}

The background radiation produced by atomic processes in the $^{76}$Ge target
was obtained from a GEANT4 simulation. The simulated atomic background is
compared with the response-corrected spectrum at $E_e$ = 7.8 MeV in
Fig.~\ref{fig:AtBgle} and at $E_e$ = 12.3 MeV in Fig.~\ref{fig:AtBghe}. The
atomic background amounts in average to only a few percent of the intensity in
the spectrum, but coincides with that above the neutron threshold, which proves
the right magnitude. This behavior is similar to that found in previous studies
\cite{rus09M,rus08,schw13,mas14,mak14,mas15,mak16,shi18} and shows that the
experimental spectrum contains a considerable amount of nuclear strength in a
quasicontinuum. This is formed by a large number of unresolved transitions with
small intensities that are the result of the increasing nuclear level density
at high energy in combination with the finite detector resolution. Because of
the different orders of magnitude, the nuclear intensity distribution resulting 
from the subtraction of the simulated atomic background is not very sensitive
to uncertainties of the latter, for which we assume 5\%.
The nuclear intensity distribution contains ground-state (elastic) transitions
and, in addition, branching (inelastic) transitions to lower-lying excited
states as well as transitions from those states to the ground state
(cascade transitions). The different types of transitions cannot be clearly
distinguished. However, for the determination of the photoabsorption cross
section and the partial widths $\Gamma_0$, the intensities of the ground-state 
transitions are needed.
Therefore, contributions of inelastic and cascade transitions have to be
subtracted from the spectra. We corrected the intensity distributions by
simulating $\gamma$-ray cascades from the levels in the entire energy region
using the code $\gamma$DEX \cite{sch12,mas12}. This code works analogously to
the strategy of the code DICEBOX \cite{bec98} developed for $(n,\gamma)$
reactions, but in addition it includes also the excitation from the ground
state. In the present simulations, level schemes (nuclear realizations)
including states with $J$ = 0, ..., 5 were created. Known low-lying levels were
taken into account up to about 3 MeV. Partial widths were varied in the
individual nuclear realizations applying the Porter-Thomas distribution
\cite{por56}.
Level densities were calculated by using the constant-temperature model
\cite{gil65} with the parameters $T$ = 0.92(1) MeV and $E_0$ = 0.13(5) MeV
adjusted to experimental level densities \cite{egi09}. In the individual
nuclear realizations, the values of $T$ and $E_0$ were varied randomly within a
Gaussian distribution with a standard deviation corresponding to the
uncertainties given in Ref.~\cite{egi09}. The parity distribution of the level
densities was modeled according to the information given in Ref.~\cite{alq03}.

\begin{figure}
\epsfig{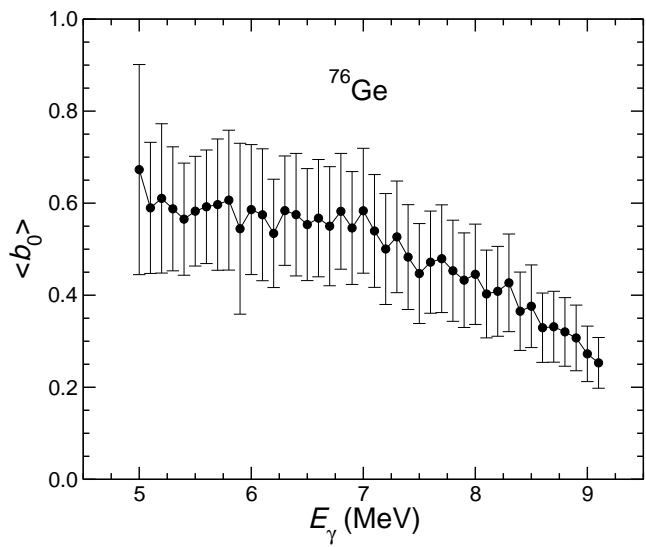} 
\caption{\label{fig:b0}Average branching ratios of ground-state transitions
  resulting from the simulations of statistical $\gamma$-ray cascades up to
  $S_n$ as described in the text.}
\end{figure}

The first inputs for the photon strength function simulations were assumed to
be Lorentz-shaped. For the $E1$ strength, a sum of three Lorentz functions (TLO)
that account for a triaxial deformation of the nucleus was used with parameters
described in Refs.~\cite{jun08,gro17}. In the present case, deformation
parameters of $\beta_2$ = 0.26 \cite{ram01} and $\gamma = 26^\circ$ \cite{del10}
were applied. The parameters for the $M1$ and $E2$ strengths were taken from
global parametrizations of $M1$ spin-flip resonances and $E2$ isoscalar
resonances, respectively \cite{cap10}. Low-lying levels were also taken into
account. Spectra of $\gamma$-ray cascades were generated for groups of levels
in energy bins of $\Delta E$ = 100 keV. Starting from the high-energy end of
the intensity distribution, that contains ground-state transitions only, the
simulated intensities of the ground-state transitions were normalized to the
experimental ones in the considered bin. The intensity distribution of the
branching transitions was subtracted from the total intensity distribution.
Applying this procedure step-by-step for each energy bin moving toward the
low-energy end of the spectrum, one obtains the intensity distribution of the
ground-state transitions. Simultaneously, the branching ratios $b_0(E)$ of the
ground-state transitions are determined for each energy bin. In an individual
nuclear realization, the branching ratio $b_0(E)$ is calculated as the ratio of
the sum of the intensities of the ground-state transitions from all levels in
$\Delta E$ to the total intensity of all transitions depopulating those levels
to either any low-lying levels or the ground state
\cite{rus09M,mas12,sch12,mas14L,schw13,mak14}. Branching ratios
$\langle b_0(E) \rangle$, averaged over the many nuclear realizations
in the present cascade simulations, are illustrated in Fig.~\ref{fig:b0}.

The uncertainty of the number of counts $N(E)$ in an energy bin of the
experimental intensity distribution was deduced as 
\begin{equation}
\label{eq:dN}
\delta N(E) = \sqrt{N(E)} 
             + \sum_{E'}{\left[\sqrt{N(E' > E)}~b(E' \rightarrow E)\right]},
\end{equation}
where $b(E'\rightarrow E)$ is the branching intensity from bin $E'$ to bin $E$.
We transform $N(E)$ to the scattering cross section according to
\begin{equation}
\label{eq:sig}
\sigma_{\gamma\gamma}(E) = N(E)/
\left[\epsilon(E)~\Phi_\gamma(E)~W(E)~N_N~\Delta t~\Delta E~\right]
\end{equation}
with the quantities defined in Eq.~(\ref{eq:sigs}), the absolute detector
efficiency $\epsilon(E)$, the measuring time $\Delta t$, and the bin width
$\Delta E$. The absorption cross section in each bin is obtained as 
$\sigma_\gamma(E) = \sigma_{\gamma \gamma}(E)/b_0(E)$ for each nuclear 
realization. Finally, the absorption cross sections of each bin were obtained
by averaging over the values of the individual nuclear realizations.

\begin{figure}
\epsfig{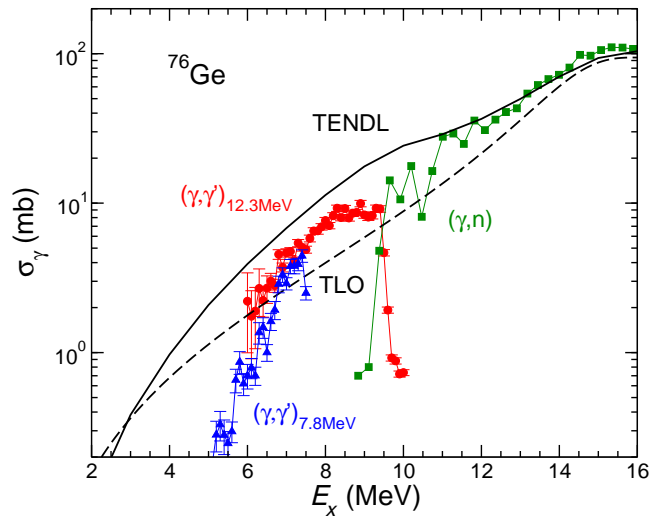} 
\caption{\label{fig:CSgggn}Photoabsorption cross sections of $^{76}$Ge
  resulting from the present $(\gamma,\gamma')$ experiments at $E_e$ = 7.8 MeV
  (blue triangles) and at $E_e$ = 12.3 MeV (red circles), from $(\gamma,n)$
  data (green squares) taken from Ref.~\cite{car76}, from calculations using
  the TALYS code as given in the TENDL-2019 library (black solid curve), and
  from the TLO with deformation parameters given in the text (black dashed
  curve).}
\end{figure}

The simulations were performed iteratively, where the strength function
obtained from an iteration step was used as the input for the next step. We
note that the simulations are little sensitive to the shape of the first input
strength function, which was tested, for example, in Refs.~\cite{sch08,sch12}.
The iteration was stopped when the input strength function and the output
strength function were in agreement within their respective uncertainties.
The cross section obtained in the last iteration step is adopted as the final
absorption cross section. Toward low energy, the uncertainties increase due 
to the use of the spectrum-stripping method and the cross sections do not
converge. Besides, the assumption of a statistical quasicontinuum becomes
invalid and individual states become important. Therefore, cross sections
cannot be determined reliably below about 6 MeV. The procedure just decsribed
was also performed for the measurement at $E_e$ = 7.8 MeV to extend the cross
section data to low excitation energy. The absorption cross sections obtained
in this way are listed in Tables~\ref{tab:CSle},~\ref{tab:CShe}, and graphed
in Fig.~\ref{fig:CSgggn}.
The uncertainties of the cross-section values include statistical uncertainties
of the spectrum, the given uncertainties of the efficiency and the subtracted
simulated background spectrum, uncertainties of the flux resulting from the
integrated cross sections of the $^{11}$B levels and the uncertainties of the
level-density parameters given in the text above. Systematic uncertainties of
level-density models can result in additional uncertainties of up to about 20\%,
which are not included here. These deviations of modeled from experimentally
determined level densities and between the various level-density models are,
for example, discussed for the case of $^{76}$Ge in Ref.~\cite{voi19}.

\begin{table}
  \caption{\label{tab:CSle}Photoabsorption cross section of $^{76}$Ge deduced
    from the present $(\gamma,\gamma')$ experiment at $E_e$ = 7.8 MeV.}
\begin{ruledtabular}
\begin{tabular}{rc} 
  $E_\gamma$ (MeV) & $\sigma$ (mb) \footnotemark[1]\\
  \hline
    5.0 &  0.18(3)  \\
    5.1 &  0.18(5)  \\
    5.2 &  0.28(7)  \\
    5.3 &  0.33(7)  \\
    5.4 &  0.28(7)  \\
    5.5 &  0.25(5)  \\
    5.6 &  0.30(5)  \\
    5.7 &  0.66(12) \\
    5.8 &  0.87(15) \\
    5.9 &  0.62(10) \\
    6.0 &  0.70(13) \\
    6.1 &  0.79(12) \\
    6.2 &  0.70(10) \\
    6.3 &  1.36(22) \\
    6.4 &  1.48(23) \\
    6.5 &  1.00(13) \\
    6.6 &  1.62(21) \\
    6.7 &  1.93(27) \\
    6.8 &  2.9(4)   \\
    6.9 &  3.3(4)   \\
    7.0 &  2.9(3)   \\
    7.1 &  3.8(4)   \\
    7.2 &  3.8(4)   \\
    7.3 &  3.9(4)   \\
    7.4 &  4.4(4)   \\
    7.5 &  2.51(26) \\
\end{tabular}
\end{ruledtabular}
\footnotetext[1]{Absorption cross section resulting from the experimental
intensity distribution including the quasicontinuum, corrected for branching
intensities and branching ratios obtained from $\gamma$-ray cascade
simulations. The uncertainties include statistical uncertainties of
the spectra (see Sec.~\ref{sec:corr}), the given uncertainties of the
efficiencies and the subtracted simulated background spectra, uncertainties of
the flux resulting from the integrated cross sections of the $^{11}$B levels
and the given uncertainties of the level-density parameters.} 
\end{table}

\begin{table}
  \caption{\label{tab:CShe}Photoabsorption cross section of $^{76}$Ge deduced
    from the present $(\gamma,\gamma')$ experiment at $E_e$ = 12.3 MeV.}
\begin{ruledtabular}
\begin{tabular}{rc} 
  $E_\gamma$ (MeV) & $\sigma$ (mb) \footnotemark[1]\\
  \hline
    6.0 &  2.2(12) \\ 
    6.1 &  1.8(8)  \\
    6.2 &  1.9(8)  \\
    6.3 &  2.7(9)  \\   
    6.4 &  2.2(5)  \\
    6.5 &  2.7(6)  \\
    6.6 &  3.0(4)  \\
    6.7 &  2.8(3)  \\
    6.8 &  4.5(3)  \\
    6.9 &  3.7(3)  \\
    7.0 &  4.7(3)  \\
    7.1 &  4.8(3)  \\
    7.2 &  4.1(2)  \\
    7.3 &  5.4(3)  \\
    7.4 &  5.1(3)  \\
    7.5 &  4.9(3)  \\
    7.6 &  5.8(3)  \\
    7.7 &  6.5(3)  \\
    7.8 &  6.5(3)  \\
    7.9 &  6.9(3)  \\
    8.0 &  7.6(4)  \\
    8.1 &  7.1(3)  \\
    8.2 &  8.2(4)  \\
    8.3 &  9.2(4)  \\
    8.4 &  8.0(4)  \\
    8.5 &  9.2(4)  \\
    8.6 &  8.0(4)  \\
    8.7 &  8.6(4)  \\
    8.8 &  8.7(4)  \\
    8.9 &  9.9(5)  \\
    9.0 &  8.3(4)  \\
    9.1 &  8.1(4)  \\
    9.2 &  8.2(4)  \\
    9.3 &  9.2(5)  \\
    9.4 &  9.1(5)  \\
    9.5 &  4.7(3)  \\
    9.6 &  1.93(9) \\
    9.7 &  0.92(4) \\
    9.8 &  0.88(4) \\
    9.9 &  0.72(3) \\
   10.0 &  0.74(3) \\
\end{tabular}
\end{ruledtabular}
\footnotetext[1]{Absorption cross section resulting from the experimental
intensity distribution including the quasicontinuum, corrected for branching
intensities and branching ratios obtained from $\gamma$-ray cascade
simulations. The uncertainties include statistical uncertainties of
the spectra (see Sec.~\ref{sec:corr}), the given uncertainties of the
efficiencies and the subtracted simulated background spectra, uncertainties of
the flux resulting from the integrated cross sections of the $^{11}$B levels
and the given uncertainties of the level-density parameters.} 
\end{table}

\section{Discussion}
\label{sec:discus}

\begin{figure}
\epsfig{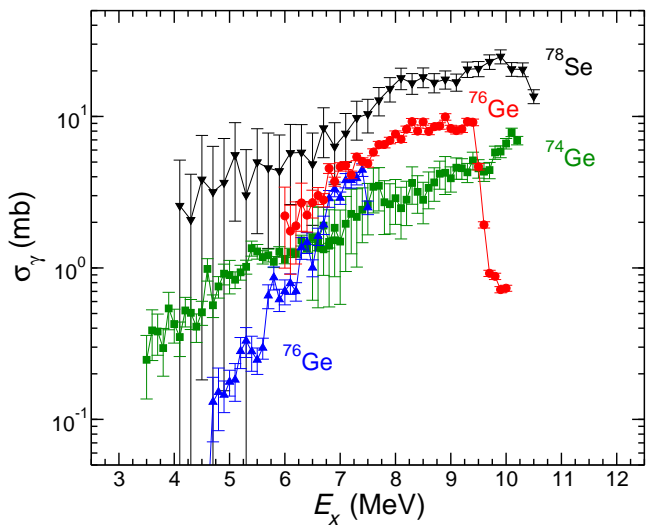} 
\caption{\label{fig:CS74Ge}Photoabsorption cross sections of $^{74}$Ge
  (green squares), $^{76}$Ge (blue triangles and red circles), and $^{78}$Se
  (black triangles), resulting from $(\gamma,\gamma')$ experiments at
  $\gamma$ELBE. The data for $^{74}$Ge were taken from Ref.~\cite{mas15} and
  the data for $^{78}$Se were taken from Refs.~\cite{sch12,mak16}.}
\end{figure}

The photoabsorption cross section resulting from the present experiments is
compared with the cross section of the $(\gamma,n)$ reaction \cite{car76} in
Fig.~\ref{fig:CSgggn}. In addition, the TLO with the deformation parameters
given in Sec.~\ref{sec:corr} and the photoabsorption cross section given in the
latest TALYS-based evaluated nuclear data library (TENDL-2019) \cite{ten19} are
displayed. In the latter, the standard Lorentzian (Brink-Axel model)
\cite{bri55,axe62,kaw20} was used as a strength function in the
$(\gamma,\gamma')$ reaction \cite{kon20}. The present $(\gamma,\gamma')$ cross
section shows extra strength above the TLO in the energy region from about
6 MeV to $S_n$, which is attributed to the PDR. The shape of the experimental
cross section is fairly well approximated by the TENDL cross section because
of its relatively smooth behavior. The cross section of $^{76}$Ge is compared
with those of the neighboring isotope $^{74}$Ge \cite{mas15} and of the isotone
$^{78}$Se \cite{sch12,mak16} resulting from analogous experiments and methods in
Fig.~\ref{fig:CS74Ge}. In the PDR region from about 6 to 9 MeV, the cross
section of $^{76}$Ge appears to be in average by a factor of about two higher
than that of $^{74}$Ge, but by a factor of about two smaller than that of
$^{78}$Se. Toward low energy, the cross section obtained from the
low-energy measurement on $^{76}$Ge drops more rapidly than the ones in
$^{74}$Ge and $^{78}$Se. These relatively large differences to nuclides with two
neutrons less or two protons more, respectively, are remarkable, and
interesting to be tested by nuclear models.

\begin{figure}
\epsfig{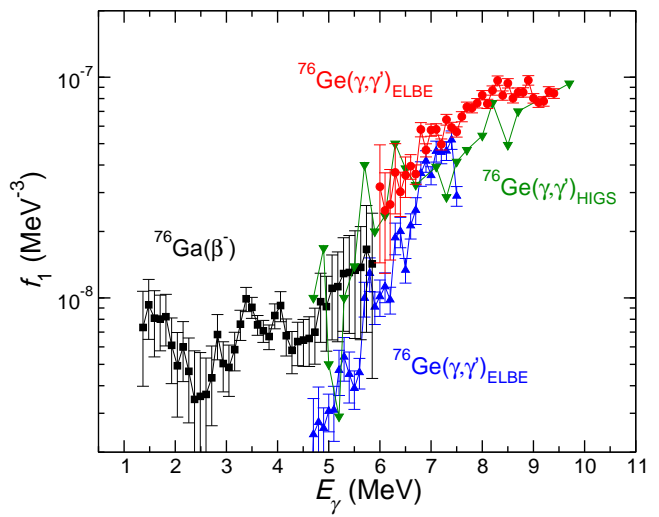} 
\caption{\label{fig:f1} Photon strength functions for $^{76}$Ge from the present
  $(\gamma,\gamma')$ experiment (blue triangles and red circles), from
  $\beta$ decay of $^{76}$Ga (black squares), taken from Ref.~\cite{spy14}, and
  from a $(\gamma,\gamma')$ experiment at HI$\gamma$S (green triangles), taken
  from Ref.~\cite{ton17}.}
\end{figure}

The photon strength function deduced from the present photoabsorption cross
section of $^{76}$Ge is compared with preliminary results from a
$(\gamma,\gamma')$ experiment with quasimonoenergetic photons at the
HI$\gamma$S facility \cite{ton17} and with data obtained from an experiment
applying the so-called Oslo method in connection with the $\beta$ decay of
$^{76}$Ga \cite{spy14} in Fig.~\ref{fig:f1}. The data from the HI$\gamma$S
and the $\beta$ decay experiments exceed the present data below 6 MeV. Between
7 and 9 MeV, the HI$\gamma$S data amount in average to about 70\% of the
present data. They were obtained from an analysis of mainly resolved elastic
transitions \cite{ton21}. This means that average branching ratios
$\langle b_0(E) \rangle$ in energy bins are overestimated at high excitation
energy and, hence, the photoabsorption cross section is underestimated.
Besides, strength in the quasicontinuum was not taken into account in that
analysis. Similar discrepancies resulting from missing strength in the
quasicontinuum were also reported in other recent studies
\cite{mak16,mue20,wil20}. On the other hand, the present
$\langle b_0(E) \rangle$ include uncertainties of the inputs for the
statistical cascade simulations, for example, of the used level-density model.

Whereas the strength functions deduced from photoabsorption cross sections
contain exclusively ground-state transitions from $J$ = 1 states, the strength
functions obtained from light-ion induced reactions or from $\beta$ decay
comprise a large number of transitions linking many states of various spins up
to about $J$ = 10, which may cause different characteristics of the strength
functions. The strength function of $^{76}$Ge from $\beta$ decay continues to
low transition energies that belong to cascade transitions between close-lying
levels at high excitation energy. It shows the characteristic upbend below
about 2.5 MeV that was also observed in the isotopic neighbors $^{73}$Ge and
$^{74}$Ge \cite{ren16}. This low-energy enhancement of strength has been
described in shell-model calculations as resulting from the large strengths of
many $M1$ transitions between states generated by recoupling the spins of
protons and neutrons in high-$j$ orbitals \cite{sch13L,bro14Fe,sie17}.
Interestingly, a bump appears between about 3 and 4.5 MeV in $^{76}$Ge in
addition to the low-energy upbend. In the shell-model calculations, such a bump
appears in open-shell nuclei and has been related to the scissors resonance,
which develops in deformed nuclides \cite{sch17}. A pronounced bump around
3 MeV in addition to the low-energy upbend was also observed in Sm isotopes
\cite{sim16,naq19}.
The $(\gamma,\gamma')$ data principally contain also inelastic and cascade
transitions between close-lying states at high excitation energy that
contribute to the low-energy enhancement \cite{sch19}. Those transitions are
hidden in the huge background in the low-energy parts of the $\gamma$-ray
spectra. An identification of the low-energy enhancement in $(\gamma,\gamma')$
experiments may be feasible by applying coincidence techniques.

\section{Summary}

The dipole-strength distribution in $^{76}$Ge was studied up to the
neutron-separation energy in photon-scattering experiments at the $\gamma$ELBE
bremsstrahlung facility using two electron energies. A total of 210 levels was
identified. Spins $J$ = 1 were deduced from angular correlations of
ground-state transitions. A $\gamma$ transition in the region of interest for
the $0\nu\beta\beta$ decay has not been observed. 

The intensity distribution obtained from the measured spectra was corrected for
the detector response and a simulated spectrum of photons scattered from the
target by atomic interactions was subtracted. The remaining spectrum contains
a continuum part in addition to the resolved peaks, which was included in the
determination of the photoabsorption cross section. An assignment of inelastic
transitions to particular levels and, thus, the determination of branching
ratios was, in general, not possible. Therefore, we performed simulations of
statistical $\gamma$-ray cascades to estimate intensities of branching
transitions. These were subtracted from the experimental intensity distribution
and the remaining intensities of ground-state transitions were corrected on
average for their branching ratios. In this way, a continuous photoabsorption
cross section was derived for the energy range from about 5 MeV up to the
neutron threshold at 9.4 MeV.

The absorption cross section of $^{76}$Ge displays an extra strength on top of
the tail of a Lorentz function for the GDR in the range between 6 and 9 MeV
that can be considered as the PDR. The shape of the PDR is relatively smooth
and approximated by cross sections calculated in the statistical model as given
in the TENDL library. The PDR is more pronounced and by a factor of about
two higher in magnitude than in the isotope $^{74}$Ge, but on the other hand by
a factor of about two smaller than in the isotone $^{78}$Se. The strength
function of $^{76}$Ge resulting from the present work is comparable with the
ones from other experiments in the PDR region, but drops rapidly toward small
energy.

\section{Acknowledgments}

We thank the operating crew of the ELBE accelerator for the cooperation.
R.~M. acknowledges support by the U.S. Department of Energy, Office of Science,
Office of Nuclear Physics under grant No. LANLEM78.

\end{document}